\documentclass[12pt]{article}
\pdfoutput=1
\usepackage{tikz}
\usepackage{putex}
\usepackage{feyn}
\usepackage[vcentermath]{youngtab}
\usepackage{subfig}
\usepackage{lscape}
\usepackage{graphicx}
\usepackage{epstopdf}
\usepackage{enumerate}
\usepackage{cite}
\usepackage{tensor} 
\usepackage{slashed}
\usepackage{amsmath}
\usepackage{amssymb}
\usepackage{mathrsfs}
\usepackage{lgrind}
\usepackage{youngtab}

\usepackage{amsmath,amssymb,braket}

\usepackage{bbm}
\usepackage{hyperref}

\numberwithin{equation}{section}

\newcommand {\be} {\begin {equation}}
\newcommand {\ee} {\end {equation}}

\newcommand {\bes} {\begin {equation*}}
\newcommand {\ees} {\end {equation*}}

\def\CH{{\cal H}}

\newcommand{\beq}{\begin{equation}}
\newcommand{\eeq}{\end{equation}}

\def\be{ \begin{equation} }
\def\ee{ \end{equation} }

\def \be {\beta}

\def \beq { \begin{equation}}
\def \eeq {\end{equation}}

\begin{document}

\preprint{PUPT-2569}

\institution{PU}{Department of Physics, Princeton University, Princeton, NJ 08544}
\institution{PCTS}{Princeton Center for Theoretical Science, Princeton University, Princeton, NJ 08544}
\institution{HU}{Department of Physics, Harvard University, Cambridge, MA 02138}

\title{Spectrum of Majorana Quantum Mechanics\\
with $O(4)^3$ Symmetry
}

\authors{Kiryl Pakrouski\worksat{\PU}, Igor R.~Klebanov\worksat{\PU,\PCTS}, Fedor Popov\worksat{\PU},
Grigory Tarnopolsky\worksat{\HU}
}

\abstract{
We study the quantum mechanics of 3-index Majorana fermions $\psi^{abc}$ governed by a quartic Hamiltonian with $O(N)^3$ symmetry.
Similarly to the Sachdev-Ye-Kitaev model, this tensor model has a solvable large $N$ limit dominated by the melonic diagrams.  
For $N=4$ the total number of states is $2^{32}$, but they naturally break up into distinct sectors according to the charges under the $U(1)\times U(1)$ 
Cartan subgroup of one of the
$O(4)$ groups. The biggest sector has vanishing charges and contains  over $165$ million states. Using a Lanczos algorithm, we determine the 
spectrum of the low-lying states in this and other sectors. We find that the absolute ground state
is non-degenerate. If the $SO(4)^3$ symmetry is gauged, it is known from earlier work that the model has $36$ states and
a residual discrete symmetry. We study the discrete symmetry group
in detail; it gives rise to degeneracies of some of the gauge singlet energies.
We find all the gauge singlet energies numerically and use the results to propose exact analytic expressions for them.
}

\date{}

\maketitle

\tableofcontents

%%%%%%%%%%%%%           1              %%%%%%%%%%%%%%
\section{Introduction}

In recent literature there has been considerable interest in the quantum mechanical models where the degrees of freedom are fermionic tensors of rank 3 or higher
\cite{Witten:2016iux, Klebanov:2016xxf}. Similarly to the Sachdev-Ye-Kitaev model \cite{Sachdev:1992fk,Kitaev:2015,Kitaev:2017awl},
these models have solvable large $N$ limits dominated by the so-called melonic diagrams 
\cite{Gurau:2009tw,Bonzom:2011zz,Carrozza:2015adg}. In this limit they become solvable with the use of  
Schwinger-Dyson equations as were derived earlier for the SYK-like models \cite{Kitaev:2015,Polchinski:2016xgd,Maldacena:2016hyu,Jevicki:2016bwu,Gross:2016kjj,
Klebanov:2016xxf,Kitaev:2017awl}.
While this spectrum of eigenstates is discrete and bounded for finite $N$, the low-lying states become dense for large $N$ leading
to the (nearly) conformal behavior where it makes sense to calculate the operator scaling dimensions. In the SYK model, the number of states is $2^{N_{\rm SYK}/2}$, and
numerical calculations of spectra
have been carried out for rather large values of $N_{\rm SYK}$ \cite{Garcia-Garcia:2016mno,Cotler:2016fpe,Gur-Ari:2018okm}. 
They reveal a smooth distribution of energy eigenvalues, which typically has no degeneracies and is almost symmetric under $E\rightarrow -E$.

The corresponding studies of spectra in the tensor models of \cite{Witten:2016iux} and \cite{Klebanov:2016xxf} have been carried out in 
\cite{Krishnan:2016bvg,Klebanov:2017,Krishnan:2017ztz,Chaudhuri:2017vrv,Krishnan:2017txw, Krishnan:2017lra, Krishnan:2018hhu,Klebanov:2018nfp}, 
but in these cases the numerical limitations have been more severe -- the number of states 
grows as $2^{N^3/2}$ in the $O(N)^3$ symmetric model of \cite{Klebanov:2016xxf} and as $2^{2 N^3}$ in the $O(N)^6$ symmetric 
Gurau-Witten (GW) model \cite{Witten:2016iux}. 
The results have shown an interesting structure. For example, for the $N=2$ GW model the exact values of the 140 $SO(2)^6$ invariant 
energies were found \cite{Krishnan:2018hhu}. Due to the discrete
symmetries, there are only 5 distinct $E<0$ eigenvalues and each one squares to an integer (the singlet spectrum also contains 50 zero-energy states).

The $O(N)^3$ model \cite{Klebanov:2016xxf}, 
has the Hamiltonian \footnote{
Compared to \cite{Klebanov:2016xxf,Klebanov:2018nfp} we have set the overall dimensionful normalization constant $g$ to $4$ in order to simplify the equations.}
\begin{align}
&H = \psi^{abc}\psi^{ab'c'} \psi^{a'bc'}\psi^{a'b'c} -  \frac{1 } {4} N^4\ ,\\
& \{ \psi^{abc}, \psi^{a'b'c'}\} = \delta^{aa'}\delta^{bb'}\delta^{cc'}\ , \qquad a,b,c=0, 1, \ldots N-1\ . 
\label{Htraceless}
\end{align}
For $N=2$ there are only two gauge singlet states with $E=\pm 8$. For $N=3$, as for any odd $N$,
there are none  \cite{Klebanov:2018nfp}.
While the complete spectra of (\ref{Htraceless})
can be calculated for $N=2$ and $3$ using a laptop, this is no longer true for $N=4$, where the total number of states
is $2^{32}$. However, they split into smaller sectors according to the charges $(Q_0, Q_1)$ of the $U(1)\times U(1)$ Cartan subgroup of one of the $SO(4)$ groups.
The most complicated and interesting is the $(0,0)$ sector; it is the part of the 32 qubit spectrum at the "half-half-filling,"
i.e. where the first 16 qubits contain 8 zeros and 8 ones, and the same applies to the remaining 16 qubits.
In particular, all the $SO(4)^3$ invariant states are in this subsector; their number, $36$, was found using the gauged version of
the free fermion theory \cite{Klebanov:2018nfp}.
 Since there are over $165$ million states at half-half-filling, the spectrum cannot be determined completely.
However, using a Lanczos algorithm, we will be able to determine a number of low-lying eigenstates. We will also be able to find the complete spectrum of the
$36$ gauge singlet states, including their transformation properties under the residual discrete symmetries of the model where the $SO(4)^3$ symmetry is gauged.
Thus,
our work reveals the spectrum of a finite-$N$ system without disorder, which is nearly conformal and solvable in the large-$N$ limit, 
and identifies the discrete symmetries crucial for efficient numerical studies of such finite systems. 

Using our numerical results we are able to infer the exact expressions for all the singlet eigenvalues. In particular, the ground state energy,\footnote{
For some results on the ground states in the SYK and related models see \cite{Garcia-Garcia:2016mno,Cotler:2016fpe,Iyoda:2018osm}.
} 
which is numerically 
$E_0\approx -160.140170$, agrees well with
$E_0 = - \sqrt{32 \left (447+  \sqrt{125601} \right )}$.
Other gauge singlet energies either have similar expressions or are simply square roots of integers.
This suggests that the Hamiltonian can be diagonalized exactly 
analytically.

\section{Discrete symmetries acting on the gauge singlets}
\label{basicsetup}

For any even $N$, 
if we gauge the $SO(N)^3$ symmetry, there remain some gauge singlet states \cite{Klebanov:2018nfp}, which are annihilated by the
symmetry charges
\begin{equation}
Q_{1}^{aa'}= \frac {i}{2} [\psi^{abc},\psi^{a'bc}]\ , \qquad
Q_{2}^{b b'}= \frac {i}{2} [\psi^{ab c},\psi^{a b' c}]\ , \qquad
Q_{3}^{c c'}= \frac {i}{2} [\psi^{ab c},\psi^{a b c'}]\ .
\label{SONcharges}
\end{equation}
These states may still have degeneracies due to the residual discrete symmetries. 
Indeed,
each $O(N)$ group contains a $Z_2$ parity symmetry which is an axis reflection. 
For example, inside $O(N)_1$ there is parity symmetry  $P_1$ which send 
$\psi^{0 b c} \rightarrow - \psi^{0 b c}$ for all $b,c$ and leaves all
other components invariant.
The corresponding generator is
\begin{equation}
P_1= P_1^\dagger = 2^{N^2/2} \prod_{bc} \psi^{0 b c}\ .
\label{parityone}
\end{equation}
One can indeed check that
\begin{equation}
P_1 \psi^{abc} P_1^\dagger = (-1)^{\delta_{a,0}+N^2}  \psi^{abc}\ .
\end{equation}
Similarly, there are $Z_2$ generators $P_2$ and $P_3$ inside $O(N)_2$ and $O(N)_3$.

It is also useful to introduce unitary operators $P_{ij}$ associated with permutations of the $O(N)_i$ and $O(N)_j$ groups:
\begin{align}
&P_{23} = P_{23}^\dagger= i^{n(n-1)/2} \prod_a \prod_{b>c} (\psi^{abc}- \psi^{acb} )\ , \notag \\
& P_{12} = P_{12}^\dagger= i^{n(n-1)/2} \prod_c \prod_{a>b} (\psi^{abc}- \psi^{bac} )\  , 
\end{align}
where $n=N^2 (N-1)/2$ is the number of fields in the product.
They satisfy 
\begin{align}
P_{23} \psi^{abc} P_{23}^\dagger =(-1)^{N^2 (N-1)/2} \psi^{acb} \ , \qquad 
P_{12} \psi^{abc} P_{12}^\dagger =(-1)^{N^2 (N-1)/2} \psi^{bac}\ .
\end{align}
These permutations flip the sign of $H$ \cite{Bulycheva:2017ilt,Klebanov:2018nfp}:
\begin{align}
P_{23} H P_{23}^\dagger =-H\ , \qquad   P_{12} H P_{12}^\dagger =-H \ . 
\end{align}
This explains why the spectrum is symmetric under $E\rightarrow -E$.

We now define the cyclic permutation operator
$P=P_{12} P_{23}$ such that
\begin{align}
P \psi^{abc} P^\dagger = \psi^{cab}\ , 
\qquad P H P^\dagger = H\ , \qquad P^3=I\ .
\end{align}
Thus, $P$ is the generator of the $Z_3$ symmetry of the Hamiltonian.
Applying the $Z_3$ symmetry to the parity reflections $P_i$ we see that
\begin{align}
P P_1 P^\dagger = P_2\ , \qquad P P_2 P^\dagger = P_3\ , \qquad P P_3 P^\dagger = P_1\ .
%\qquad P^\dagger P_1 P= P_3\ . 
\label{consistentparity}
\end{align}
%Therefore, the three parity operators belong to the same conjugacy class.
 
Forming all the possible products of $I, P, P_1, P_2, P_3$, 
we find that the full discrete symmetry group contains 24 elements.
Using the explicit representation (\ref{parityone}) for $P_1$, and the analogous ones for $P_2$ and $P_3$, we note that the three parity operators commute with each other.
Furthermore,
\begin{align}
[\Pi, P]=0\ , \qquad \Pi = P_1 P_2 P_3\ , \qquad \Pi^2=I\ .
\end{align}
Therefore, $\Pi$ commutes with all the group elements, so that the group has a $Z_2$ factor with elements $I$ and $\Pi$.
The symmetry group turns out to be
 $A_4\times Z_2$, and the 12 elements of the alternating group $A_4$ are 
\begin{align}
I\ , P_1\ , P_2\ , P_1 P_2\ , P\ , P^2\ , P_1P\ , P_2 P\ , P_1 P_2 P\ , P_1 P^2\ , P_2 P^2\ , P_1 P_2 P^2\ .
\end{align}
Each of them can be associated with a sign preserving permutation of 4 ordered elements, and the action is
\begin{align}
& P_1 (a_0, a_1, a_2, a_3) = (a_1, a_0, a_3, a_2)\ ,\notag \\
& P_2 (a_0, a_1, a_2, a_3) = (a_2, a_3, a_0, a_1)\ ,\notag \\
& P_3 (a_0, a_1, a_2, a_3) = (a_3, a_2, a_1, a_0)\ ,\notag \\
& P (a_0, a_1, a_2, a_3) = (a_0, a_3, a_1, a_2)\ .
\label{afour}
\end{align}

The degenerate $SO(N)^3$ invariant states of a given non-zero energy form irreducible representations of $A_4\times Z_2$. 
For even $N$ we can choose a basis where 
all the wavefunctions and 
matrix elements of the Hamiltonian are real. In this case we should study the representation of the symmetry group over the field $\mathbb{R}$.
The degrees of the irreducible representations of $A_4$ over that field are $1,2,3$.
The $Z_2$ factor does not change the degrees since both irreducible representations of $Z_2$,
the trivial one and the sign one, have degree $1$. 

Let us discuss the representations of $A_4$ in more detail. Using a reference eigenstate $|\psi_0\rangle$ not invariant under the $Z_3$ subgroup $I, P, P^2$, 
we can form a triplet of states 
\begin{align}
|\psi_0\rangle\ , \qquad  P |\psi_0\rangle\ ,\qquad P^2 |\psi_0\rangle\ .
\label{triplet}
\end{align}
If the parities $(P_1, P_2, P_3)$ of the state $|\psi_0\rangle$ are the same, 
then we can form a linear combination which transforms trivially under the $Z_3$,
\begin{align}
|\psi\rangle\ = \frac{1} {\sqrt 3} (1+ P+ P^2) 
|\psi_0\rangle\ , \qquad  P |\psi\rangle\ = |\psi\rangle\ ,
\end{align}
while the remaining 2 linear combination form the degree 2 representation of $Z_3$,
\begin{align}
P |\psi_1\rangle = |\psi_2 \rangle \ , \qquad P |\psi_2\rangle = - |\psi_1\rangle  - |\psi_2\rangle\, ,
\end{align}
where $\ket{\psi_1} = \frac{1}{\sqrt{3}}\ket{\psi_0} - \ket{\psi}$.
Because of this, some eigenstates have degeneracy 2.

If the parities $(P_1, P_2, P_3)$ of the state $|\psi_0\rangle$ are not equal, then the triplet representation (\ref{triplet}) of the full discrete group is irreducible. 
For example for $(P_1, P_2, P_3)= (+,+,-)$, i.e.
\begin{align}
P_1 |\psi_0\rangle= |\psi_0\rangle\ , \qquad  P_2 |\psi_0\rangle= |\psi_0\rangle\ ,  \qquad P_3 |\psi_0\rangle= -|\psi_0\rangle\ , 
\label{parfirst}
\end{align}
we find that the parities of the states $P |\psi_0\rangle$ and  $P^2 |\psi_0\rangle$ are
given by the cyclic permutations of $(+,+,-)$. Indeed, using (\ref{consistentparity}), we find that the parities of the state $P |\psi_0\rangle$ are
\begin{align}
P_1 P |\psi_0\rangle= - P |\psi_0\rangle\ , \qquad  P_2 P |\psi_0\rangle= P |\psi_0\rangle\ ,  \qquad P_3 P |\psi_0\rangle= P |\psi_0\rangle\ . 
\label{second}
\end{align}
% and analogously for the state $P^2 |\psi_0\rangle$.
Thus,
each of the states in the triplet (\ref{triplet}) has a distinct set of parities. Then it is impossible to form linear combinations which are eigenstates of the parities, and
we have an irreducible representation of $A_4$ of degree $3$.
In this situation we find that an energy eigenvalue has degeneracy $3$. 

We also note the relations
\begin{align}
 P_{23} P_1 P_{23}^\dagger = (-1)^{N (N^2-1)/2} P_1\ , 
\quad 
P_{12} P_1 P_{12}^\dagger = (-1)^{N (N^2-1)/2} P_2\ , \quad
P_{13} P_1 P_{12}^\dagger = (-1)^{N (N^2-1)/2} P_3\ , 
\label{parityflips}
\end{align}
and their cyclic permutations.
Since an operator $P_{ij}$ maps an eigenstate of energy $E$ into an eigenstate of energy $-E$, 
we see that such mirror states have the same parities when $N/2$ is even, but opposite parities when $N/2$ is odd.
%Our explicit results for parities in the $N=2$ and $N=4$ cases will be consistent with this rule.  

%\subsection{The zero-energy sector}

For the states at zero energy, the discrete symmetry group is enhanced to $48$ elements because the permutation generators $P_{ij}$ map them into themselves. 
%Thus, for the zero-energy states the discrete group is enhanced and contains $48$ elements.
Using the relations (\ref{parityflips}) we find
\begin{align}
P_{12}\Pi P_{12}^\dagger = (-1)^{N (N^2-1)/2} \Pi \ ,
\end{align}
which implies that $\Pi=P_1 P_2 P_3$ commutes or anti-commutes with other elements depending on the value of $N$.
Focusing on the case where $N(N^2-1)/2$ is even and the sign above is positive (this includes $N=4$ which is our main interest in this paper), we find 
that $\Pi$ commutes with all other generators, 
so that the group has a $Z_2$ factor with elements $I$ and $\Pi$.
The symmetry group for $E=0$ turns out to be
 $S_4\times Z_2$, which is the full cube group. Its subgroup $S_4$ is formed out of the products of $I, P_1, P_2, P_{12}, P_{23}, P_{13}$. 
The parity generators are realized in the same way as in (\ref{afour}), while the permutations act by the natural embedding $S_3 \subset S_4$:
\begin{align}
& P_{12} (a_0, a_1, a_2, a_3) = (a_0, a_2, a_1, a_3)\ ,\notag \\
& P_{23} (a_0, a_1, a_2, a_3) = (a_0, a_1, a_3, a_2)\ ,\notag \\
& P_{13} (a_0, a_1, a_2, a_3) = (a_0, a_3, a_2, a_1)\ .
\label{sfour}
\end{align}
The degrees of the irreducible representations of $S_4$ are $1,1,2,3,3$.

\section{Diagonalization of the Hamiltonian}

%\section{A basis for operators and states}
%\label{somebounds}

The Majorana fermions $\psi^{abc}$ may be thought of as generators of the Clifford algebra in $N^3$-dimensional Euclidean space. 
Restricting to the cases where $N$ is even, the dimension of the Hilbert space is $2^{N^3/2}$, and the states may be represented by series
of $N^3/2$ ``qubits" $ |s\rangle $, where $s=0$ or $1$.
It is convenient to introduce operators \cite{Krishnan:2017txw,Klebanov:2018nfp}
\begin{gather}
	\bar{c}_{abk}= \frac{1}{\sqrt 2} \left (\psi^{ab(2k)}+ i \psi^{ab (2k+1)} \right ), \quad  c_{abk} = \frac{1}{\sqrt 2} \left (\psi^{ab(2k)}- i \psi^{ab (2k+1)} \right )\,, \notag\\
	\left\{c_{abk}, c_{a' b' k'} \right\}=\left\{\bar{c}_{abk}, \bar{c}_{a' b' k'} \right\}=0,\quad \left\{\bar{c}_{abk}, c_{a'b' k'}\right\}= \delta_{a a'} \delta_{b b'} \delta_{k k'}\,, 
\label{ladderoperator}
	\end{gather}
where $ a,b=0, 1, \ldots, N-1,$ and $ k=0, \ldots, \frac{1} {2} N -1$. 
In this basis the $O(N)^2\times U(N/2)$ symmetry is manifest, and
the Hamiltonian is  \cite{Krishnan:2017txw,Klebanov:2018nfp}
\begin{gather}
H = 2\big(\bar{c}_{abk}\bar{c}_{ab'k'}c_{a'bk'}c_{a'b'k}-\bar{c}_{abk}\bar{c}_{a'bk'}c_{ab'k'}c_{a'b'k}\big)
 \,.
\label{onHamilt}
\end{gather}
If we number the qubits from $0$ to $\frac{1} {2} N^3 -1$, then operators $c_{abk}, \bar c_{abk}$ correspond to qubit number $N^2 k+N b+a$. 
%For $N=4$, if we number the qubits from $0$ to $31$, then the operators $c_{abk}, \bar c_{abk}$ correspond to qubit number $16k+4b+a$. 

In the basis (\ref{ladderoperator}) the parity operators $P_i$ corresponding to $i$-th group $O(N)$ are 
\begin{align}
P_1 = \prod_{b=0}^{N-1}  \prod_{k=0}^{N/2-1}  [\bar{c}_{0bk}, c_{0b k}], \quad 
 P_2 = \prod_{a=0}^{N-1}  \prod_{k=0}^{N/2 -1 } [\bar{c}_{a0k}, c_{a0 k}], \quad
 P_3 = \prod_{a=0}^{N-1}  \prod_{b=0}^{N-1}   (\bar{c}_{ab0} + c_{ab 0})   . 
\label{parityops}
\end{align} 
The operator $P_3$ implements charge conjugation on the $k=0$ operators, i.e. it acts to interchange  $\bar{c}_{ab0}$ and $c_{ab 0}$.
%the particle-hole conjugation on the first $N^2$ qubits, which are acted on by operators $\bar{c}_{ab0}, c_{ab 0}$. 
This conjugation is a symmetry of $H$. In fact, for each $k$ the Hamiltonian is symmetric under the interchange of $\bar{c}_{abk}$ and ${c}_{abk}$.

The $U(1)^{N/2}$ subgroup of the $U(N/2)$ symmetry is realized simply. The corresponding charges,
\begin{align}
Q_k = \sum_{a,b} \frac {1} {2} [\bar{c}_{abk}, c_{ab k}] \ , \qquad k=0, \ldots, \frac{1} {2} N -1\ ,
\end{align}
are the Dynkin lables of a state of the third $SO(N)$ group,
and the spectrum separates into sectors according to their values.
The oscillator vacuum state satisfies
\begin{align}
c_{ab k}\ket{\rm vac} = 0\ , \qquad 
%q_{ab k}\ket{vac} = -\frac{1}{2} \ket{vac}\ , \qquad 
Q_k \ket{\rm vac} = -\frac{N^2}{2} \ket{\rm vac}\ ,
\end{align}
and other states are obtained by acting on it with some number of $\bar c_{ab k}$.

%\section{Diagonalization of the Hamiltonian}

For $N=4$ 
the total number of states is $2^{32}= 4294967296$, but they break up into $17^2=289$ smaller sectors due to the conservation of
the $U(1)\times U(1)$ charges $Q_0$ and $Q_1$. The biggest sector is $(Q_0, Q_1)= (0,0)$; it consists of  
$\frac{(16 !)^2}{ (8 !)^4}= 165636900$ states. The next biggest are the $4$ sectors $(\pm 1, 0)$ and $(0,\pm 1)$; each of them 
contains $147232800$ states.
The smallest $4$ sectors are $(\pm 8, \pm 8)$, and each one consists of just $1$ state; each of these states has $E=0$.
In general, the spectrum in the $(q,q')$ sector is the same as in $(q',q)$ due to the symmetry of $H$ under interchange of the $c_{ab0}$ and $c_{ab1}$
oscillators.

Let us first study the $(0,0)$ sector. 
These states are obtained by acting on $\ket{vac}$ with $8$ raising operators $\bar{c}_{ab0}$ 
and $8$ raising operators $\bar{c}_{ab1}$. In the qubit notation, both the first $16$ qubits, and the second $16$ qubits, have equal number, $8$, of zeros and ones.
Clearly, all the $SO(4)^3$ invariant states are in this sector.\footnote{
There are additional constraints on the gauge singlet wave functions, but we will not discuss them explicitly here.}
While the numbers of such ``half-half-filled" states is still very large, they turn out to be tractable numerically because the matrix we need to diagonalize is rather sparse.
This has allowed us to study the low-lying eigenvalues of $H$, which occur in various representations of $SO(4)^3$.
To find the gauge singlet energies, we study the operator proportional to $H+ 100 \sum_{i=1}^3 C_2^i$, where the quadratic Casimir 
of the $SO(N)_1$ symmetry is $C_2^1=\frac{1}{2} Q_{1}^{a a'}Q_{1}^{aa'}$,
%\begin{equation}
%C_2^1=\frac{1}{2} Q_{1}^{a a'}Q_{1}^{aa'}\ ,
%\end{equation}
and analogously for $SO(N)_2$ and $SO(N)_3$.
The Lanczos algorithm allows us to identify the lowest eigenvalues of this operator, which all correspond to $SO(4)^3$ invariant states; the non-singlets receive large additive contributions
due to the second term.

 \begin{figure}[h!]
                \centering
                \includegraphics[width=10cm]{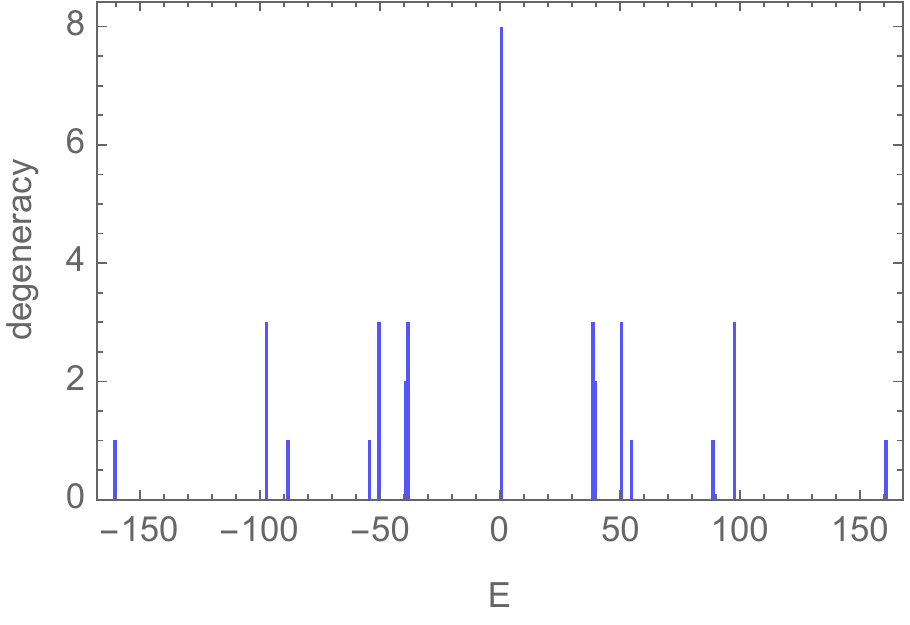}
                \caption{Spectrum of gauge singlets in the $O(4)^3$ model}
                \label{o4o4o4sp}
\end{figure} 

\begin{table}[!h!]
\caption{\label{tab:singlets444} The list of all the $SO(4)^3$ invariant states including their parities $P_i$.}
\begin{center}
\begin{tabular}{|c|c|c|c|c|c|c|c|}
\hline
$E$ 			& $P_1$  	& $P_2$ 	& $P_3$ 	& $E$ 		& $P_1$  	& $P_2$ 	& $P_3$\\
\hline
$-160.140170$	& 1		& 1		&1 		& 160.140170 	& 1		& 1		&1\\
\hline
$-97.019491$	& 1		& 1		&$-1$		&   97.019491  	& 1		& 1		&$-1$\\
$-97.019491$	& $- 1$		& 1		&1		&   97.019491  	& $- 1$		& 1		&1\\
$-97.019491$	& 1		& $-1$		&1		&   97.019491  	& 1		& $-1$		&1\\
\hline
$-88.724292$	& $-1$		& $-1$		&$-1$		&   88.724292	& $-1$		& $-1$		&$-1$		 \\
\hline
$-54.434603$	& 1		& 1		&1 		&   54.434603	& 1		& 1		&1         \\
\hline
$-50.549167$	& 1 		&   1 		& $-1$		&   50.549167	& 1 		& 1 		& $-1$		 \\
$-50.549167$	& $-1$ 		& 1 		&  1		&   50.549167	& $-1$ 		& 1 		& 1		 \\
$-50.549167$	& 1 		& $-1$ 		& 1		&   50.549167	& 1 		& $-1$ 		& 1		 \\
\hline
$-39.191836$	& 1		& 1		& 1		&   39.191836	& 1		& 1		& 1		 \\
$-39.191836$	& 1		& 1		& 1		&   39.191836	& 1		& 1		& 1	 \\
\hline
$-38.366652$	& 1 		& $-1$ 		& $-1$		&   38.366652	& 1 		& $-1$ 		& $-1$	 \\
$-38.366652$	& $-1$ 		& 1 		& $-1$		&   38.366652	& $-1$ 		& 1 		& $-1$	 \\
$-38.366652$	& $-1$ 		& $-1$ 		& 1		&   38.366652	& $-1$ 		& $-1$ 		& 1	 \\
\hline
0.000000		& 1		& 1		&1 		&      0.000000	 & $-1$		& $-1$		& $-1$\\
0.000000		& $-1$		& 1		&1 		&      0.000000	 & 1		& $-1$		& $-1$\\
0.000000		& 1		& $-1$		&1 		&      0.000000	 & $-1$		& 1		&$-1$\\
0.000000		& 1		& 1		&$-1$ 		&      0.000000	 & $-1$		& $-1$		&1\\
\hline
\end{tabular}
\end{center}
\end{table}

In table \ref{tab:singlets444} we list the energies and parities of all 36 $SO(4)^3$ invariant states. 
In order to identify the values of $P_i$, we calculated the low-lying spectrum of operator 
\begin{equation} H+ 100 \sum_{i=1}^3 C_2^i+ \sum_{i=1}^3 a_i P_i\ ,
\end{equation} 
where $a_i$ are
unequal small coefficients.\footnote{The states at $\pm 39.191836$ are doubly degenerate and have identical parities; these states form the degree $2$ representation of the $Z_3$
subgroup of $A_4$.
To split such double degeneracies we added a small amount of noise to the Hamiltonian.}
The biggest degeneracy is found for the $E=0$ states; it corresponds to the $2^3$ independent choices of the three parities. 
Since the discrete group acting on the $E=0$ states is $S_4\times Z_2$, which is the full cube group, we find two different irreducible representations of
$S_4$: the trivial one of degree $1$
and the standard one of degree $3$. We may associate the eight $E=0$ states with the vertices of a cube. 
The energies of the gauge singlet states and their degeneracies are 
plotted in
Figure \ref{o4o4o4sp}. 

Some of the energies agree within the available precision with square roots of integers: 
$8\sqrt{23} \approx 38.366652$, $8\sqrt{24} \approx 39.191836 $, and $8\sqrt{123} \approx 88.724292 $. 
Furthermore, the $4$ eigenvalues with parities $(1,1,1)$, $\pm 160.140170$ and  $\pm 54.434603$,  
are approximations to the analytic expressions  $\pm \sqrt{32 \left (447\pm \sqrt{125601} \right )}$, 
%$\sqrt{ 14304\pm 32 \sqrt{125601}}$,
while the triplet eigenvalues, $\pm 97.019491$ and  $\pm 50.549167$, are approximations to
$\pm \sqrt{32 \left (187\pm \sqrt{11481} \right )}$. 
To demystify these exact results, we note that there are only two $SO(4)^3$ invariant states with
$P_1= P_2= P_3=-1$ (see Table \ref{tab:singlets444}).
%parities $(-1,-1,-1)$.
Since the Hamiltonian has symmetry under $E\rightarrow -E$, the eigenvalue equation in this subsector must have the form of the second order even polynomial: $E^2 - A_1=0$. This explains why some of the eigenvalues are simply square roots. On the other hand, there are four  $SO(4)^3$ and $Z_3$ invariant states with $P_1= P_2= P_3=1$. Thus, the eigenvalue equation in this subsector must have the form
\begin{equation}
E^4 + 2 A_2 E^2 + A_3=0 \ ,
\end{equation}
and this explains why some of the energies satisfy $E^2=- A_2\pm \sqrt{A_2^2- A_3}$. Similar symmetry considerations explain the form of all the gauge singlet energies in terms of the square roots. We leave exact derivation of the parameters $A_i$ for future work.
%$\sqrt{ 5984\pm 32\sqrt{11481}}$. 
%These expressions in terms of square roots
%suggest that there is an exact solution for the singlet spectrum.
%\footnote{
%}

\begin{table}[!h!]
\caption{\label{tab:numerics444HOnly} The low-lying energies in the $(0,0)$ sector, i.e. at half-half filling, including the values of the quadratic Casimirs of each $SO(N)$ group.
When the $C_2^i$ are not all equal, there are additional states of the same energy with their values obtained by a cyclic permutation.
}
\begin{center}
\begin{tabular}{|c|c|c|c|c|c|c|}
\hline
$C_2^1$ 	& $C_2^2$ 		&  $C_2^3$  		& $E$\\
\hline
0	&	0  		& 0 				& -160.140170  \\
0	&	4 		& 8 				& -136.559039  \\
0	&	0 		& 12 				& -136.417554	\\
0	&	0 		& 24 				& -128.490197	\\
4	&	4 		& 4	 			& -122.553686	\\
0	&	0 		& 12 				& -121.606040	\\
4	&	8		& 8 				& -121.552284	\\
4	&	8		& 8 				& -120.699077	\\
4	&	8		& 8 				& -119.685636	\\
0	&	8		& 12 				& -119.659802	\\
0	&	12		& 8 				& -119.204505	\\
0	&	8		& 4 				& -118.699780	\\
0	&	4		& 16				& -118.541049	\\
4	&	4 		& 4 				& -116.774758	\\
\hline
\end{tabular}
\end{center}
\end{table}

The list of all the low-lying energy levels in the $(0,0)$ sector, singlets and non-singlets, and the corresponding values of quadratic Casimirs $C_2^i$, 
is shown in table \ref{tab:numerics444HOnly}.
In order to identify the values of $C_2^i$, we have calculated the low-lying spectrum of 
$H+\sum_{i=1}^3 a_i C_2^i$ where $a_i$ are
unequal small coefficients. 
When the $C_2^i$ are not all equal, there are also states of the same energy with their values obtained by a cyclic permutation.
For example, at $E= -136.559039$ we find states with $(C_2^1, C_2^2, C_2^3)= (0,4,8), (4,8,0), (8,0,4)$.

\begin{table}[!h!]
\caption{\label{tab:numerics444HOnlyPM1Hole} The low-lying states in the sectors $(\pm 1, 0)$ and $(0,\pm 1)$, i.e. with one extra hole (h) or particle (p) added to half-half-filling
The energies are the same within the accuracy shown, which is a good test of our diagonalization procedure. When the $C_2^i$ are not all equal, 
there are additional states of the same energy with their values obtained by a cyclic permutation.}
\begin{center}
\begin{tabular}{|c|c|c|c|c|c|c|c|}
\hline
$C_2^1$ 	& $C_2^2$ 		&  $C_2^3$  		& $E_h=E_p$			\\
\hline
3	&	3  		& 3 			& -140.743885  	\\
3	&	3 		& 9	 		& -128.059272 		\\
3	&	3 		& 15	 		& -124.547555		\\
3	&	9 		& 9 			& -118.371087		\\
3	&	3 		& 9			& -117.798571		\\
3	&	3 		& 19 			& -115.861910		\\
3	&	9 		& 9 			& -114.885221		\\
3	&	3 		& 15 			& -114.660576		\\
3	&	3 		& 9 			& -114.539928		\\
\hline
\end{tabular}
\end{center}
\end{table}

Absent from the list in table \ref{tab:numerics444HOnly} is the lowest possible value of the quadratic Casimir, $C_2=3$, which corresponds to the 
$(1/2,0)+(0,1/2)$ irrep, i.e.
fundamental representation $4$ of $SO(4)$.
Let us proceed to the sectors adjacent to one-particle and one-hole sectors, $(\pm 1, 0)$ and $(0,\pm 1)$,
which contain some of the additional representations, including the 
$(4,4,4)$ of $SO(4)^3$. The refined bound \cite{Klebanov:2018nfp} for this representation gives
$| E_{(4,4,4)}| < 72 \sqrt 5 \approx 160.997$,
while the actual lowest state in this representation has $E \approx -140.743885$. The low-lying states in the sectors
$(\pm 1, 0)$ and $(0,\pm 1)$ are given in table~\ref{tab:numerics444HOnlyPM1Hole}. 
We have also calculated the energies in other charge sectors. We find that the absolute ground state lies in the $(0,0)$ sector:
as the magnitudes of charges increase, the energies tend to get closer to $0$.

\section{Supplementary Material for "Spectrum of Majorana Quantum Mechanics with $O(4)^3$ Symmetry"}

For $N=4$ we can use the explicit representation in terms of direct products of $32$ $2\times 2$ matrices:
\begin{align}
\label{Majoranabasis}
& \sqrt 2 \psi^{000}=X\otimes \mathbbm{1} \otimes \mathbbm{1}\ldots \otimes \mathbbm{1}\,, \qquad  
\sqrt 2 \psi^{001}=Y\otimes \mathbbm{1} \otimes \mathbbm{1} \ldots \otimes \mathbbm{1}\,, \notag\\
& \sqrt 2  \psi^{100}=Z\otimes X \otimes \mathbbm{1} \ldots \otimes \mathbbm{1}\,, \qquad 
\sqrt 2 \psi^{101}=Z\otimes Y \otimes \mathbbm{1} \ldots \otimes \mathbbm{1}\,, \notag\\
%& \sqrt 2  \psi^{200}=Z\otimes Z\otimes X \otimes \mathbbm{1} \ldots \otimes \mathbbm{1}\,, \qquad 
%\sqrt 2 \psi^{201}=Z\otimes Z\otimes Y \otimes \mathbbm{1} \ldots \otimes \mathbbm{1}\,, \notag\\
& \ldots \notag\\
%& \sqrt 2  \psi^{330}=Z\otimes X \otimes \mathbbm{1} \ldots \otimes \mathbbm{1}\,, \qquad 
%\sqrt 2 \psi^{331}=Z\otimes Y \otimes \mathbbm{1} \ldots \otimes \mathbbm{1}\,, \notag\\
%& \sqrt 2  \psi^{002}=Z\otimes X \otimes \mathbbm{1} \ldots \otimes \mathbbm{1}\,, \qquad 
%\sqrt 2 \psi^{003}=Z\otimes Y \otimes \mathbbm{1} \ldots \otimes \mathbbm{1}\,, \notag\\
& \sqrt 2  \psi^{232}=Z\ldots \otimes Z  \otimes X \otimes \mathbbm{1}\,, \qquad
\sqrt 2 \psi^{233}=Z\ldots \otimes Z\otimes Y \otimes \mathbbm{1} \,,\notag\\
& \sqrt 2 \psi^{332}=Z\ldots \otimes Z\otimes Z\otimes X\,, \qquad
\sqrt 2 \psi^{333}=Z\ldots \otimes Z\otimes Z\otimes Y\,,
\end{align}
where $X,Y,Z$ stand for the Pauli matrices $\sigma_x, \sigma_y, \sigma_z$. Their action on a qubit is
\begin{align}
&X|0\rangle =|1\rangle, \quad  Y|0\rangle =-i|1\rangle, \quad  Z|0\rangle =-|0\rangle\,, \notag\\
&X|1\rangle =|0\rangle, \quad  Y|1\rangle =i|0\rangle, ~~~\quad  Z|1\rangle =|1\rangle\,. \label{actst} 
\end{align}

The Hamiltonian %(\ref{onHamilt}) 
becomes
\begin{align}
H &= 2\big(\bar{c}_{ab0}\bar{c}_{ab'0}c_{a'b0}c_{a'b'0}-\bar{c}_{ab0}\bar{c}_{a'b0}c_{ab'0}c_{a'b'0}\big)
 + 2\big(\bar{c}_{ab1}\bar{c}_{ab'1}c_{a'b1}c_{a'b'1}-\bar{c}_{ab1}\bar{c}_{a'b1}c_{ab'1}c_{a'b'1}\big) \notag \\
& + 4\big(\bar{c}_{ab0}\bar{c}_{ab'1}c_{a'b1}c_{a'b'0}-\bar{c}_{ab0}\bar{c}_{a'b1}c_{ab'1}c_{a'b'0}\big)
 \, ,
\label{onHamiltnew}
\end{align}
where in the first line we find two copies of the Hamiltonian of the $O(4)^2\times O(2)$ model, which was solved in \cite{Klebanov:2018nfp}. 
Each of these systems contains $16$ qubits, and the second line creates
a coupling between the two systems.

The expressions for the parity operators are, omitting the direct product signs,
\begin{align}
&P_1= Z111Z111Z111Z111Z111Z111Z111Z111\ , \notag \\
&P_2= ZZZZ111111111111ZZZZ111111111111\ , \notag \\
  &P_3= YXYXYXYXYXYXYXYX1111111111111111 \ .
\end{align}
The operator $P_3$ implements, up to a sign, the particle-hole conjugation on the first $16$ qubits. 
These parity operators may be used only on the $SO(4)^3$ invariant states. For example, a rotated form of 
$P_3= 2^{8} \prod_{ab} \psi^{a b 0}$ is  
\begin{align}
\tilde P_3= 
2^{8} \prod_{ab} \psi^{a b 1}= XYXYXYXYXYXYXYXY1111111111111111\ .
\end{align}
It has the same eigenvalues as $P_3$ on the singlets because 
\begin{equation}
\tilde P_3 P_3= ZZZZZZZZZZZZZZZZ1111111111111111\ ,
\end{equation}
 which is equal to $1$ when acting
on the singlet states, where the first 16 qubits are half-filled.

The ground state energy we find
is close to the lower bound \cite{Klebanov:2018nfp}
$E_{bound}= -\frac{1 }{4}N^3(N+2) \sqrt{N-1}\approx -166.277 $.
The ratio of $E_{0}$ and  $E_{bound}$ can be calculated in the large $N$ limit using the exact propagator, and it is found to be 
$\approx 0.41$ \cite{Klebanov:2018nfp}. Since we find $E_0/E_{bound} \approx 0.96$, 
this suggests that large $N$ approximations cannot be applied for $N=4$.

In some charge sectors it is not hard to determine the energy spectrum.
For example, in the $(- 7, - 7)$ sector, which contains $256$ states 
$ \bar{c}_{ab0}\bar{c}_{a'b'1} |{\rm vac}\rangle$, only the second line in the Hamiltonian (\ref{onHamiltnew}) acts non-trivially, and we find $E=\pm 16$
with multiplicity $15$, and $E=0$ with multiplicity $226$. Due to the conjugation symmetry %, noted below (\ref{parityops}), 
the same spectrum is found in the $(-7,7)$ sector.
In each of the $(\pm 6, \pm 6)$ sectors, some of the energies are square roots of integers, including the ground state $E=-24\sqrt{5}$. 
Finally, let us note that in each sector of the form $(q, \pm 8)$ or $(\pm 8, q)$ the Hamiltonian is isomorphic to that of the $O(4)^2\times O(2)$ model, which was solved in 
\cite{Klebanov:2018nfp}, and therefore has the same integer spectrum. The lowest and highest 
energies, occurring for $q=0$, are $\pm 64$. 

From the value of quadratic Casimir we may infer
the $(j_1, j_2)$ representation of $SO(4)\sim SU(2)\times SU(2)$ using the formula
\begin{equation}
C_2 (j_1, j_2)= 2 \big (j_1(j_1+1) +  j_2(j_2+1)\big )\ .
\end{equation}
For example, $C_2=4$ corresponds to the $(1,0)+(0,1)$ irrep of dimension 6; $C_2=8$ corresponds to the $(1,1)$ irrep of dimension $9$;
$C_2=12$ corresponds to the $(2,0)+(0,2)$ irrep of dimension 10; etc.

\section*{Acknowledgments}

The simulations presented in this article were performed on computational resources managed and supported by Princeton's Institute for Computational Science $\&$ Engineering and OIT Research Computing.
We are grateful to Yakov Kononov and Douglas Stanford  for useful discussions.
KP was supported by the Swiss National Science Foundation through the Early Postdoc.Mobility grant P2EZP2$\_$172168.
The work of IRK and FP was supported in part by the US NSF under Grant No.~PHY-1620059. 
The work of GT was supported in part by  the MURI grant W911NF-14-1-0003 from ARO and by DOE grant de-sc0007870.

%%%%%%%%%%%%%
 
\bibliographystyle{ssg}
\bibliography{Spectrum}

\begingroup\raggedright\begin{thebibliography}{10}

\bibitem{Witten:2016iux}
E.~Witten, ``{An SYK-Like Model Without Disorder},''
  \href{http://xxx.lanl.gov/abs/1610.09758}{{\tt 1610.09758}}.

\bibitem{Klebanov:2016xxf}
I.~R. Klebanov and G.~Tarnopolsky, ``{Uncolored random tensors, melon diagrams,
  and the Sachdev-Ye-Kitaev models},'' {\em Phys. Rev.} {\bf D95} (2017), no.~4
  046004, \href{http://xxx.lanl.gov/abs/1611.08915}{{\tt 1611.08915}}.

\bibitem{Sachdev:1992fk}
S.~Sachdev and J.~Ye, ``{Gapless spin fluid ground state in a random, quantum
  Heisenberg magnet},'' {\em Phys. Rev. Lett.} {\bf 70} (1993) 3339,
  \href{http://xxx.lanl.gov/abs/cond-mat/9212030}{{\tt cond-mat/9212030}}.

\bibitem{Kitaev:2015}
A.~Kitaev, ``{A simple model of quantum holography},''.
  \url{http://online.kitp.ucsb.edu/online/entangled15/kitaev/},\url{http://online.kitp.ucsb.edu/online/entangled15/kitaev2/}.
  Talks at KITP, April 7, 2015 and May 27, 2015.

\bibitem{Kitaev:2017awl}
A.~Kitaev and S.~J. Suh, ``{The soft mode in the Sachdev-Ye-Kitaev model and
  its gravity dual},'' {\em JHEP} {\bf 05} (2018) 183,
  \href{http://xxx.lanl.gov/abs/1711.08467}{{\tt 1711.08467}}.

\bibitem{Gurau:2009tw}
R.~Gurau, ``{Colored Group Field Theory},'' {\em Commun. Math. Phys.} {\bf 304}
  (2011) 69--93, \href{http://xxx.lanl.gov/abs/0907.2582}{{\tt 0907.2582}}.

\bibitem{Bonzom:2011zz}
V.~Bonzom, R.~Gurau, A.~Riello, and V.~Rivasseau, ``{Critical behavior of
  colored tensor models in the large N limit},'' {\em Nucl. Phys.} {\bf B853}
  (2011) 174--195, \href{http://xxx.lanl.gov/abs/1105.3122}{{\tt 1105.3122}}.

\bibitem{Carrozza:2015adg}
S.~Carrozza and A.~Tanasa, ``{$O(N)$ Random Tensor Models},'' {\em Lett. Math.
  Phys.} {\bf 106} (2016), no.~11 1531--1559,
  \href{http://xxx.lanl.gov/abs/1512.06718}{{\tt 1512.06718}}.

\bibitem{Polchinski:2016xgd}
J.~Polchinski and V.~Rosenhaus, ``{The Spectrum in the Sachdev-Ye-Kitaev
  Model},'' {\em JHEP} {\bf 04} (2016) 001,
  \href{http://xxx.lanl.gov/abs/1601.06768}{{\tt 1601.06768}}.

\bibitem{Maldacena:2016hyu}
J.~Maldacena and D.~Stanford, ``{Comments on the Sachdev-Ye-Kitaev model},''
  {\em Phys. Rev.} {\bf D94} (2016), no.~10 106002,
  \href{http://xxx.lanl.gov/abs/1604.07818}{{\tt 1604.07818}}.

\bibitem{Jevicki:2016bwu}
A.~Jevicki, K.~Suzuki, and J.~Yoon, ``{Bi-Local Holography in the SYK Model},''
  {\em JHEP} {\bf 07} (2016) 007,
  \href{http://xxx.lanl.gov/abs/1603.06246}{{\tt 1603.06246}}.

\bibitem{Gross:2016kjj}
D.~J. Gross and V.~Rosenhaus, ``{A Generalization of Sachdev-Ye-Kitaev},''
  \href{http://xxx.lanl.gov/abs/1610.01569}{{\tt 1610.01569}}.

\bibitem{Garcia-Garcia:2016mno}
A.~M. Garcia-Garcia and J.~J.~M. Verbaarschot, ``{Spectral and thermodynamic
  properties of the Sachdev-Ye-Kitaev model},'' {\em Phys. Rev.} {\bf D94}
  (2016), no.~12 126010, \href{http://xxx.lanl.gov/abs/1610.03816}{{\tt
  1610.03816}}.

\bibitem{Cotler:2016fpe}
J.~S. Cotler, G.~Gur-Ari, M.~Hanada, J.~Polchinski, P.~Saad, S.~H. Shenker,
  D.~Stanford, A.~Streicher, and M.~Tezuka, ``{Black Holes and Random
  Matrices},'' {\em JHEP} {\bf 05} (2017) 118,
  \href{http://xxx.lanl.gov/abs/1611.04650}{{\tt 1611.04650}}.

\bibitem{Gur-Ari:2018okm}
G.~Gur-Ari, R.~Mahajan, and A.~Vaezi, ``{Does the SYK model have a spin glass
  phase?},'' \href{http://xxx.lanl.gov/abs/1806.10145}{{\tt 1806.10145}}.

\bibitem{Krishnan:2016bvg}
C.~Krishnan, S.~Sanyal, and P.~N. Bala~Subramanian, ``{Quantum Chaos and
  Holographic Tensor Models},'' {\em JHEP} {\bf 03} (2017) 056,
  \href{http://xxx.lanl.gov/abs/1612.06330}{{\tt 1612.06330}}.

\bibitem{Klebanov:2017}
I.~Klebanov, D.~Roberts, D.~Stanford, and G.~Tarnopolsky, ``{unpublished
  (December 2017)},''.

\bibitem{Krishnan:2017ztz}
C.~Krishnan, K.~V.~P. Kumar, and S.~Sanyal, ``{Random Matrices and Holographic
  Tensor Models},'' {\em JHEP} {\bf 06} (2017) 036,
  \href{http://xxx.lanl.gov/abs/1703.08155}{{\tt 1703.08155}}.

\bibitem{Chaudhuri:2017vrv}
S.~Chaudhuri, V.~I. Giraldo-Rivera, A.~Joseph, R.~Loganayagam, and J.~Yoon,
  ``{Abelian Tensor Models on the Lattice},''
  \href{http://xxx.lanl.gov/abs/1705.01930}{{\tt 1705.01930}}.

\bibitem{Krishnan:2017txw}
C.~Krishnan and K.~V.~P. Kumar, ``{Towards a Finite-$N$ Hologram},'' {\em JHEP}
  {\bf 10} (2017) 099, \href{http://xxx.lanl.gov/abs/1706.05364}{{\tt
  1706.05364}}.

\bibitem{Krishnan:2017lra}
C.~Krishnan, K.~V. Pavan~Kumar, and D.~Rosa, ``{Contrasting SYK-like Models},''
  {\em JHEP} {\bf 01} (2018) 064,
  \href{http://xxx.lanl.gov/abs/1709.06498}{{\tt 1709.06498}}.

\bibitem{Krishnan:2018hhu}
C.~Krishnan and K.~V. Pavan~Kumar, ``{Exact Solution of a Strongly Coupled
  Gauge Theory in 0+1 Dimensions},'' {\em Phys. Rev. Lett.} {\bf 120} (2018),
  no.~20 201603, \href{http://xxx.lanl.gov/abs/1802.02502}{{\tt 1802.02502}}.

\bibitem{Klebanov:2018nfp}
I.~R. Klebanov, A.~Milekhin, F.~Popov, and G.~Tarnopolsky, ``{Spectra of
  eigenstates in fermionic tensor quantum mechanics},'' {\em Phys. Rev.} {\bf
  D97} (2018), no.~10 106023, \href{http://xxx.lanl.gov/abs/1802.10263}{{\tt
  1802.10263}}.

\bibitem{Iyoda:2018osm}
E.~Iyoda, H.~Katsura, and T.~Sagawa, ``{Effective dimension, level statistics,
  and integrability of Sachdev-Ye-Kitaev-like models},''
  \href{http://xxx.lanl.gov/abs/1806.10405}{{\tt 1806.10405}}.

\bibitem{Bulycheva:2017ilt}
K.~Bulycheva, I.~R. Klebanov, A.~Milekhin, and G.~Tarnopolsky, ``{Spectra of
  Operators in Large $N$ Tensor Models},'' {\em Phys. Rev.} {\bf D97} (2018),
  no.~2 026016, \href{http://xxx.lanl.gov/abs/1707.09347}{{\tt 1707.09347}}.

\end{thebibliography}\endgroup

\end{document}